\documentclass[usenatbib]{mn2e}
\usepackage{times,graphics,amssymb,epsfig,subfigure,color}

\newcommand{\ber}{\begin{eqnarray}}
\newcommand{\eer}{\end{eqnarray}}

\def\labequn #1{\label{eq:#1}}
\def\labfig #1{\label{fig:#1}}
\def\labsecn #1{\label{sec:#1}}
\def\labsubsecn #1{\label{subsecn:#1}}
\def\labsubsubsecn #1{\label{subsubsecn:#1}}
\def\labtablem #1{\label{tab:#1}}

\def\equn #1{Equation~\ref{eq:#1}}

\def\fig #1{Figure~\ref{fig:#1}}
\def\secn #1{Section~\ref{sec:#1}}

\def\subsecn #1{Section~\ref{subsecn:#1}}
\def\subsubsecn #1{Section~\ref{subsubsecn:#1}}
\def\tablem #1{Table~\ref{tab:#1}}

\def\etal{et al.\ }

\def\unit #1{\,{\rm #1}}

\def\ev{\unit{eV}}
\def\kev{\unit{keV}}
\def\kms{\unit{km\,s^{-1}}}

\title[BELR as a function of black hole mass]
{A transition mass for black holes to show broad emission lines}
\author[Chakravorty \etal]
{
Susmita Chakravorty$^{1,2}$
Martin Elvis$^{2*}$,
Gary Ferland$^{3*}$ \\
\footnotesize \it $^{1}$Harvard University, Department of Astronomy, schakravorty@head.cfa.harvard.edu; \\ 
\it $^{2}$Harvard-Smithsonian Center for Astrophysics, 60 Garden Street, Cambridge, MA 02138, USA \\
\it $^{3}$Department of Physics and Astronomy, University of Kentucky, Lexington, KY 40506.
}

\begin{document}
\maketitle


\begin{abstract}
Although the super-massive (AGN) and stellar mass (XRBs) black holes have many
properties in common, the broad emission lines (BELs) are exclusively
signatures of the AGN. Based on the detection of these lines from SDSS data
bases, there seems to be no AGN with mass $M_{BH} \lesssim 10^5 M_{\odot}$. In
this paper we investigate if such low mass black holes are really non-existent
or they are undetected because the BELs in them are not produced efficiently.
Using the ionizing spectral energy distribution for a wide range of black hole
mass, $10 - 10^9 M_{\odot}$, spanning XRBs to AGN, we calculate the equivalent
widths (EWs) of ultraviolet and optical lines $\rm{Ly\alpha \,\, 1216 \, \AA}$,
$\rm{ H\beta \,\, 4861 \, \AA}$, $\rm{CIV \,\, 1549 \,\AA}$ and $\rm{MgII \,\,
2798 \, \AA}$. The LOC (locally optimally emitting cloud) model has been used
to describe the broad emission line region (BELR) for the calculations. We find
that the hardening of the SED shape with decreasing mass do not decrease the BEL
EWs. However, finite size of the BELR, as measured by the line widths, which is
controlled by the mass of the black hole, regulates the production of these
emission lines. There seems to be a peak in the EWs of the emission lines for
typical AGN black holes of $\sim 10^8 M_{\odot}$, below which the lines become
intrinsically fainter with a sharp fall-off below $\sim 10^6 M_{\odot}$. This
may be the cause of the absence of low mass AGN in SDSS.  
\end{abstract}




\section{Introduction}
\labsecn{sec:introduction}

Active galactic nuclei (AGN) and X-ray binaries (XRBs) share many properties,
but broad emission lines (BELs) are not among them. The absence of BELs in XRB
spectra are often ascribed to the harder spectrum that is emitted by the
accretion disk around the $\lesssim 10^6$ times smaller black holes. Here we
investigate the predicted BEL equivalent widths (EWs) over a wide range of
black hole masses, $10 - 10^9 M_{\odot}$ to see if there is a threshold mass
below which BELs are not expected. The possibility of a threshold mass for
strong BEL production has become interesting from recent observational results,
which seem to suggest that there are no broad line emitting black holes below
$M_{BH} \lesssim 10^5 M_{\odot}$ \citep{greene04}. Is this because there are no
such black holes, or is it that, even when accreting at substantial rates, such
black holes cannot produce BELs with detectable EWs?

Most existing photoionization calculations for the broad emission lines in AGN,
study the line strengths as a function of the overall shape of the ionizing
continuum \citep[][]{osterbrock06,leighly07} and the luminosity of the AGN.
However, mostly, these studies do not take the further step of directly
relating the emission line properties to $M_{BH}$ and $\dot{m}$. On the other
hand, dynamical measurements of $M_{BH}$, from the widths of the broad lines,
and/or using the reverberation mapping techniques,
\citep{peterson93,netzer97,kaspi00} consider an average $R_{BELR} \sim
L_{AGN}^{1/2}$ relationship and do not need to delve into the details of the
line strengths and ratios. In this study we shall use the LOC model, where the
broad lines are due to locally, optimally emitting clouds, first suggested by
\citet{baldwin95}. We calculate the line strengths of the strongest broad
lines, $\rm{Ly\alpha \,\, 1216 \, \AA}$, $\rm{ H\beta \,\, 4861 \, \AA}$,
$\rm{CIV \,\, 1549 \,\AA}$ and $\rm{MgII \,\, 2798 \, \AA}$, as a function of
the black hole mass ranging from $10 - 10^9 M_{\odot}$. 


\section{Theory}
\labsecn{sec:theory}

\subsection{The accretion disk spectral energy distribution}
\labsubsecn{subsec:SED}

The radiation from the thin accretion disc ($\alpha$-disk) may be modeled as
a sum of local blackbodies emitted from the different annuli of the disc at
different radii \citep[][hereafter
SS73]{shakura73}. The temperature of the annulus at radius $R$ is
\begin{equation} 
T(R) = 6.3 \times 10^5 \left(\frac{\dot{m}} {\dot{m}_{\rm Edd}} \right)^{\frac{1}{4}} \left(\frac {M_{BH}} {10^8M_{\odot}} \right)^{-\frac{1}{4}} \left(\frac{R}{R_s}\right)^{-\frac{3}{4}}\rm{K} 
\label{eqn:DbbTemp}
\end{equation}
\citep{peterson97,frank02} where $\dot{m}$ is the accretion
rate of the central black hole of mass $M_{BH}$, $\dot{m}_{\rm Edd}$ is its
Eddington accretion rate and $R_s = 2GM_{BH}/c^2$ is the Schwarzschild radius
($G$ is the gravitational constant and $c$ is the velocity light). The
normalisation constant $A_{dbb}$ for this spectral component is given by
\begin{equation}
A_{dbb} = \left\{ \frac {R_{in}/\rm{km}} {D/(10\,\,\rm{kpc})} \right\}^2 \cos\theta
\labequn{DbbNorm}
\end{equation}
for an observer at a distance $D$ whose line-of-sight makes and angle $\theta$ to
the normal to the disc plane. $R_{in}$ is the radial distance of the innermost
stable annulus of the accretion disc from the black hole. $R_{in}$ is assumed
to be a $3~R_s$ (as conventional).
 
\begin{figure}
\begin{center}
\includegraphics[scale = 1, width = 9 cm, trim = 0 60 0 0, clip, angle = 0]{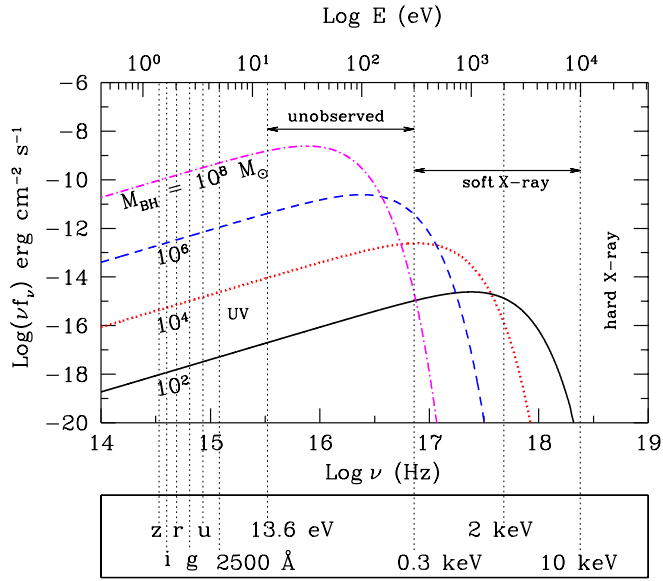}
\caption{The spectral energy distribution from accretion disks around black
holes of mass $M_{BH}\,=\,10^2, 10^4, 10^6 \,\,\rm{and}\,\,10^8\,M_{\odot}$
accreting at 0.1$\dot{m}_{\rm Edd}$ (or emitting at 0.1$L_{\rm Edd}$). The
absolute scale of the y axis is arbitrary, assuming that all the sources are at
a distance of 100 Mpc. However, the relative normalisations of the SEDs are
according to their mass ratios.  We have labeled some of the important energy
values (accompanied by the dotted vertical lines) relevant for AGN SEDs. In
particular, note the positions of the u (3551 \AA), g(4686 \AA), r(6165 \AA),
i(7481 \AA) and z(8931 \AA) filters used by the Sloan Digital Sky Survey.  }  
\labfig{Seds}
\end{center}
\end{figure}

A standard model for the spectral component from the accretion disk is
available as disk blackbody \citep[][]{mitsuda84, makishima86} in
XSPEC\footnote{http://heasarc.gsfc.nasa.gov/docs/xanadu/xspec/}
\citep{arnaud96}. We have used version 11.3 of XSPEC to generate the {\it disk
blackbody} spectral component $f_{dbb}(\nu)$ for $M_{BH} = 10 - 10^9 M_{\odot}$
at steps of $\log\,(M_{BH}/M_{\odot})\,=\,0.5$ and
$\dot{m}\,=\,0.1\,\dot{m}_{\rm Edd}$ and $\theta\,=\,30^{\circ}$ are held
constant. The spectral energy distributions (SEDs) for $M_{BH}\,=\,10^2, 10^4,
10^6 \,\,\rm{and}\,\,10^8\,M_{\odot}$ are shown in \fig{Seds} to demonstrate
how the peak of the SED and its normalisation change as a function of the black
hole mass.  The SEDs for AGN with $M_{BH} \geq 10^4\,M_{\odot}$ peak in an
energy range which is unobservable for extragalactic sources due to Galactic
extinction. As a result, for AGN, determination of the mass of the black hole
often depends on the study of the emission lines, particularly the BELs
\citep{netzer87, korista97b, kaspi00, bentz06}.

AGN are often selected from large databases like Sloan Digital Sky Survey
(hereafter SDSS) based on their broadband optical colours, which depend on the
intrinsic SED of the AGN. From \fig{Seds} we see that all of the five SDSS
filters (u-3551, g-4686, r-6165, i-7481 and z-8931 \AA) are positioned on the
linear, same slope `low energy' tail of the intrinsic AGN SEDs so that they
cannot be distinguished as black holes of different mass, merely from studying
the SDSS colours. SDSS, however observes the redshifted and not the intrinsic
SEDs of the quasars. Even when seeing the redshifted SED, there is no effect of
the mass on the SDSS colors for $z < 8.60$ for $M_{BH} \leq 10^6 M_{\odot}$  


Even if the colour selection algorithms of SDSS are not biased against
detecting the lower mass black holes, they will still drop out because the intrinsic
luminosity of the AGN decreases with the mass of the black hole. For example,
based on the flux limits of the SDSS filters \citep{stoughton02}, the $M_{BH} =
10^6 M_{\odot}$ black holes drop out beyond a redshift $\mathit{z}\geq 0.075$,
whereas a $M_{BH} = 10^8 M_{\odot}$ black hole can be seen for $\mathit{z}
\leq 1.25$. Note that the aforementioned redshift limits are based on the
continuum flux only. AGN are however detected primarily through the detection
of their emission lines. The limits based on the emission lines would be
different and we shall discuss this further in \secn{calculations}.


\subsection{The LOC model for the BELR} 
\labsubsecn{subsec:LOC}

\begin{figure*}
\begin{center}
\includegraphics[scale = 1, width = 18 cm, trim = 85 445 50 25, clip, angle = 0]{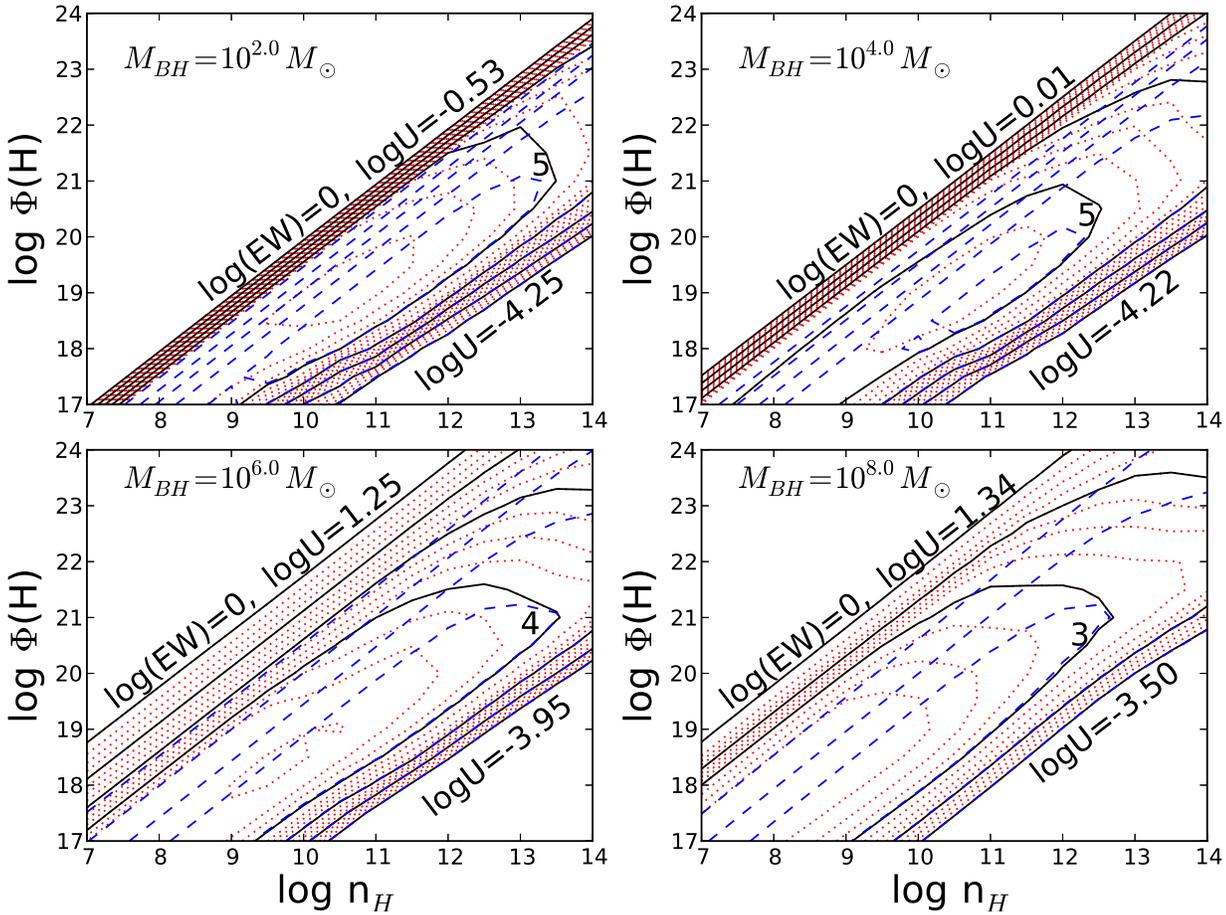}
\caption{Contour plots for the CIV (1549 \AA) equivalent width as a function of
the black hole mass in the $n_H - \Phi(H)$ plane. The bold solid black lines
are contours at steps of 1 dex and the intermediate dotted red lines are
contours at steps of 0.25 dex, for $\log N_H = 23$. In each panel, the
outermost contours correspond to EW = 1 (log(EW) = 0), whereas the innermost
contours of the maximum EW are different for different SEDs. Note that diagonal
lines with a slope of 45$^{\circ}$ in the $\log \Phi(H) - \log n_H$ plane are
lines of constant ionization parameter $U = \Phi(H)/cn_H$. In each panel, we
have labeled the value of $\log U$ corresponding to the outermost EW = 1
contour, both at the low ionization and high ionization end.  For comparison,
we also over plot the iso-contours for $\log N_H = 21$ (dashed blue lines) at
steps of 1 dex, the outermost contour being for log(EW) = 0. } 
\labfig{fig:BlrC4Contours}
\end{center}
\end{figure*}

\citet{baldwin95} and \citet{korista97a} show that, for a given SED, any
particular emission line is dominated by emission from a fairly narrow range of
gas density ($n_H$) and incident flux $\Phi(H)$ [$=Q(H)/4\pi R^2$, where $Q(H)$
is the number of hydrogen ionizing photons and $R$ is the distance of the cloud
from the source of the ionizing radiation]. This narrow range of the $n_H$ and
$\Phi(H)$ is different for different emission lines. However, as long as there
are enough clouds distributed over the relevant density and distance ranges,
all the observed emission lines can be formed with the observed line ratios.
This is thus, a model, which does not require any extreme fine tuning of $n_H$
and $\Phi(H)$ to produce all the observed emission lines. \citet{leighly07}
discuss both the successes and limitations of the LOC model. Among limitations
they mention that (a) some of the parameters of the model like the indices of
the density and radial distribution of the clouds cannot be physically
interpreted and (b) the model does not include some physical effects like
self-shielding.

\section{Calculations}
\labsecn{calculations}

\subsection{Unrestricted BELR}
\labsubsecn{subsec:UnrestrictedBelr}

For each model SED we use version C08.00 of CLOUDY\footnotemark
\citep{ferland98}.
\footnotetext{URL: http://www.nublado.org/ }
to calculate the emission line spectrum for a three dimensional parameter grid
of $\Phi(H)$, $n_H$ and column density ($N_H$). $\log\Phi(H)$ is varied from
17--24 in steps of 0.5, $\log n_H$ from 7--14 in steps of 0.5 and $\log N_H$
from 21--23 in steps of 0.5, assuming a Solar metallicity
\citep{allendeprieto01, allendeprieto02, holweger01, grevesse98} gas. We
calculate the equivalent widths (EWs) of the 42 more prominent quasar emission
lines \citep[see][for the entire list]{korista97a}. However, we demonstrate the
results for only four of the strongest lines ($\rm{Ly\alpha \,\, 1216 \, \AA}$,
$\rm{CIV \,\, 1549 \, \AA}$, $\rm{MgII \,\, 2798 \, \AA}$ and $\rm{ H\beta \,\,
4861 \, \AA}$) in this paper, which is adequate for the issues discussed here. 

We plot the iso-contours of the predicted EW (in $\log$) for the 1549 \AA \,
CIV line in the $\log~n_H - \log~\Phi(H)$ plane (\fig{fig:BlrC4Contours}) for
the four SEDs with $M_{BH}\,=\,10^{2.0}, 10^{4.0}, 10^{6.0}, \,\, \rm{and} \,\,
10^{8.0} M_{\odot}$. In the $\log \Phi(H) - \log n_H$ plane, diagonal lines
with a slope of 45$^{\circ}$ are lines of constant ionization parameter $U =
\Phi(H)/cn_H$, where the value of U increases from the bottom right to the top
left. We see that the contours form well-defined diagonal ridges (collection of
constant U lines) in the $\log \Phi(H) - \log n_H$ plane with significant EW
($> 1$). In each panel of \fig{fig:BlrC4Contours} we have also labeled the
limiting values of $\log U$, within which CIV line is efficiently produced with 
$1\le$EW$\le10^{5.75}$ for $M_{BH}\,=\,10^{2.0}M_{\odot}$,
$1\le$EW$\le10^{5.25}$ for $M_{BH}\,=\,10^{4.0}M_{\odot}$,
$1\le$EW$\le10^{4.75}$ for $M_{BH}\,=\,10^{6.0}M_{\odot}$, and
$1\le$EW$\le10^{3.75}$ for $M_{BH}\,=\,10^{8.0} M_{\odot}$. 

Deviations (of any contour) from the 45$^{\circ}$ lines show where thermal
heating of the gas begins to be more important than photoionization, usually at
high $n_H$ and high $\Phi(H)$ (upper right, each panel). For example, taking
the example of the $\log(EW)=4$ contour for $M_{BH}\,=\,10^{2.0} M_{\odot}$, we
see that it `turns over' from high $\log U (= -0.99, \log n_H \ge 12 \,\,
\rm{and} \,\,\log\Phi(H) \ge 21.5)$ and becomes diagonal again at a lower $\log
U (=-3.24)$. 

The contours for $\log N_H = 21 \,\, \rm{and} \,\, 23$, show that the low
ionization ridge (lower-right) of the iso-contours remain unaffected. However,
the high ionization ridge (upper-left) is pushed to lower $U$, e.g by 1.23 dex
for the $M_{BH}\,=\,10^{8.0} M_{\odot}$ SED. Thus for any given SED, lower
column density restricts efficient line production to a tighter range of $U$.

\begin{figure*}
\begin{center}
\includegraphics[scale = 0.5, width = 17 cm, trim = 10 390 0 50, clip, angle = 0]{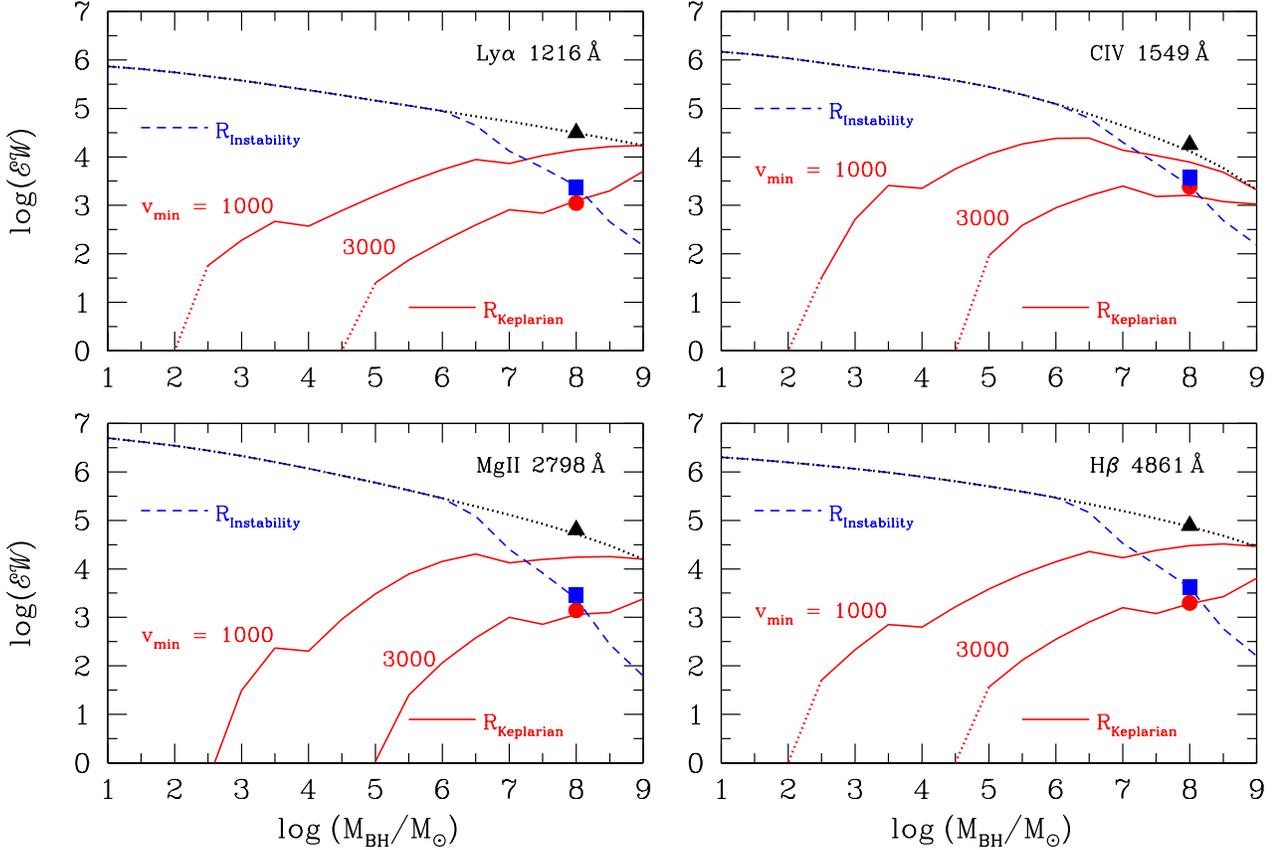}
\caption{The weighted average ($\mathcal{EW}$) for four of the strongest broad
lines, as a function of black hole mass is given by the dotted black line in
each panel. The filled black triangle at $M_{BH}\,=\,10^{8.0}M_{\odot}$ shows
the value of $\mathcal{EW}$ if we consider an SED with a disk blackbody plus a
X-ray power-law, instead of just a disk blackbody SED. The mass distribution of
$\mathcal{EW}$ is modified if a cut-off is applied on the maximum allowed
radius of the BELR; the solid red curves (with different minimum velocity of
the BELR clouds, in $\rm{km\,s^{-1}}$) if the cut-off is determined using
Keplerian mechanics (\equn{equn:Keplerian}) and the dashed blue curve if the
cut-off is determined from the considerations of the gravitational instability
radius of the accretion disk (\equn{equn:GravityUnstable}). The filled red
circle and the filled blue square denote the value of $\mathcal{EW}$ for the
disk+power-law SED for $M_{BH}\,=\,10^{8.0}M_{\odot}$ when the respective
radius cut-off schemes (as described above) are assumed. Note that for the
Keplerian cut-off mechanism, we have shown the filled circle only for the
$v_{min}=3000 \,\rm{km\,s^{-1}}$ case.} 
\labfig{AvgEwRcut}
\end{center}
\end{figure*}

To assess the contribution from all the clouds of different density and column
density and at different distances, we have to take a weighted average of EW :
\begin{equation} 
\mathcal{EW} = f \int \int \int EW(\Phi(H), n_H, N_H) \frac{d~\Phi(H)}{\Phi(H)} \frac{d~n_H}{n_H} \frac{d~N_H}{N_H}
\labequn{AvgEw}
\end{equation}
over $21 \le \log N_H \le 23$, $8 \le \log n_H \le 14$ and $18 \le \log \Phi(H)
\le 24$ all in steps of 0.5. $f = 0.2$ is the constant covering fraction adopted in this paper. $\mathcal{EW}$ (in log) as a function of $M_{BH}$
for $\rm{Ly\alpha}$, $\rm{CIV}$, $\rm{MgII}$ and $\rm{H\beta}$ is shown in
\fig{AvgEwRcut} (dotted black line).  For each of the emission lines, the
average $\mathcal{EW}$ rises monotonically with the decrease in mass. 

We further calculate the line ratios $\rm{MgII} / \rm{CIV}$ and $\rm{MgII} /
\rm{H\beta}$ and plot them against each other in \fig{LineRatios} (dotted black
line). The labeled solid black circles along the line mark the black hole
masses in units of $\log (M_{BH}/M_{\odot})$. 


\subsection{Restriction on the radius of the BELR}
\labsubsecn{subsec:RestrictedBelr}

\subsubsection{Limits from velocity of the clouds}
\labsubsubsecn{subsubsec:Kepler}

Reverberation mapping has established that the time lag between the variation
in the continuum flux and the line flux $t_{lag}$ is proportional to 1/2 power
of the full width at half maxima \citep[FWHM, see][and references
therein]{bentz06}. Hence the BELR gas seems to move with Keplerian or at least
virialized velocities, so that we can define the outermost radius of the BELR
\begin{equation}
R_{Keplerian} = G\,M_{BH}/v^2_{min}
\labequn{equn:Keplerian}
\end{equation}
corresponding to the clouds with the lowest observed velocity $v_{min}$. The
FWHM of BEL are $ \gtrsim 2000 (> 1200) \kms$ \citep{hao05}, so that $v_{min}
(= \rm{FWHM/\sqrt{8ln2}}) \gtrsim 1000 \kms$. The distribution of
$R_{Keplerian}$ are shown as solid red lines in \fig{Rmax}, for $v_{min} =
1000, 3000 \,\, \rm{and} \,\, 10000 \kms$.

\begin{table}
\begin{center}
\begin{tabular}{l l l l l l l}
\hline
\raisebox{-2.5ex}[0cm][0cm]{Cut-off} & \multicolumn{2}{c}{$\rm{CIV} - 1549 \AA$} & \multicolumn{2}{c}{$\rm{MgII} - 2798 \AA$} & \multicolumn{2}{c}{$\rm{H\beta} - 4861 \AA$} \\ \\ \cline{2-3} \cline{4-5} \cline{6-7}
used \\ ($v_{min}$ & \raisebox{-0.5ex}[0cm][0cm]{$\frac{\mathcal{EW}_{7.5}}{\mathcal{EW}_{4.0}}$} & \raisebox{-0.5ex}[0cm][0cm]{$\frac{\mathcal{EW}_{7.5}}{\mathcal{EW}_{5.0}}$} & \raisebox{-0.5ex}[0cm][0cm]{$\frac{\mathcal{EW}_{7.5}}{\mathcal{EW}_{4.0}}$} & \raisebox{-0.5ex}[0cm][0cm]{$\frac{\mathcal{EW}_{7.5}}{\mathcal{EW}_{5.0}}$} & \raisebox{-0.5ex}[0cm][0cm]{$\frac{\mathcal{EW}_{7.5}}{\mathcal{EW}_{4.0}}$} & \raisebox{-0.5ex}[0cm][0cm]{$\frac{\mathcal{EW}_{7.5}}{\mathcal{EW}_{5.0}}$} \\ 
km/s) &&&&&& \\
\hline
3000 & - & 17.52 & - & 720 & - & 32.22 \\ \\
1000 & 4.83 & 0.96 & 79.31 & 5.27 & 39.37 & 6.27 \\
\hline
\end{tabular}
\caption{The ratio of the weighted equivalent widths $\mathcal{EW}$ for different black hole masses. We have considered the finite size of the BELR, cut-off corresponding to the two Keplerian velocities $v_{min} = 3000 \,\, \rm{and} \,\, 1000 \,\, \rm{km \, s^{-1}}$.}
\labtablem{table1}
\end{center}
\end{table}

\begin{figure}
\begin{center}
\includegraphics[scale = 0.5, width = 8 cm, trim = 0 130 0 70, clip, angle = 0]{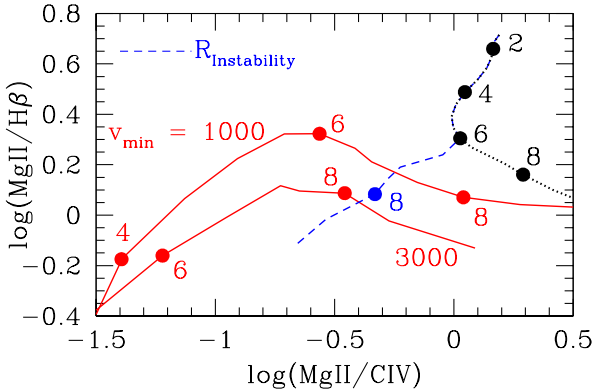}
\caption{The line ratios $\rm{MgII} / \rm{CIV}$ and $\rm{MgII} /
\rm{H\beta}$ are plotted against each other for the three different BELR. The case of Unrestricted BELR (\subsecn{subsec:UnrestrictedBelr}), with no upper limits on its size, is represented by the dotted black line. When the BELR is truncated using limits from observed Keplerian velocities (\subsecn{subsec:RestrictedBelr}) we have line ratio profiles given by the solid red lines, for the two different velocities of the clouds, namely, $v_{min}=1000\,\, \rm{and} \,\,3000 \,\rm{km\,s^{-1}}$. The dashed blue line corresponds to BELR with size limitations imposed by the gravitational instability (\subsubsecn{subsubsec:instability}). The solid circles represent the points on the line ratio profiles for different black hole masses (as labeled) in units of $\log (M_{BH}/M_{\odot})$.} 
\labfig{LineRatios}
\end{center}
\end{figure}

Introducing an $R_{Keplerian}$ cut-off in our CLOUDY simulations, produces a
pronounced cut-off in $\mathcal{EW}$ (solid red lines in \fig{AvgEwRcut}). BELs
used by SDSS to detect and identify AGN activity, $\rm{ H\beta \,\, 4861 \,
\AA}$, $\rm{CIV \,\, 1549 \,\AA}$ and $\rm{MgII \,\, 2798 \, \AA}$ are all
affected.  The ratio $\mathcal{EW}_{7.5}/\mathcal{EW}_{5.0} > 10$ for all the
emission lines, for $v_{min} = 3000 \kms$ (\tablem{table1}), where
$\mathcal{EW}_x$ is the $\mathcal{EW}$ due to the black hole mass $M_{BH}\,=\,
10^x \, M_{\odot}$. This ratio is as high as 720 for $\rm{MgII \,\, 2798 \,
\AA}$.  Below $M_{BH}\,=\, 10^{5.0} M_{\odot}$, $\mathcal{EW}$ rapidly falls
for all the lines.

The larger BELR allowed by $v_{min} = 1000 \kms$ produces a less drastic drop
and theory predicts non-zero equivalent widths upto much lower black hole
masses ($\sim 10^{2.5} M_{\odot}$).  For $v_{min} = 1000 \kms$,
$\mathcal{EW}_{7.5}/\mathcal{EW}_{5.0}$ reduces to 5.27 for $\rm{MgII \,\, 2798
\, \AA}$. \tablem{table1} list the equivalent width ratios for the different
lines. In fact for the two CIV and MgII lines the  $\mathcal{EW}$ profiles are
not monotonically decreasing functions of black hole mass, rather the peak of
the profiles shift to lower mass $\sim 10^{6.5} M_{\odot}$. 

The detection of the lines by SDSS would however, depend on the intrinsic
luminosity in the emission lines, in addition to their equivalent widths. The
intrinsic luminosity in the lines, in turn depend on the mass of the black
hole. When no limit is imposed on the outer radius of the BELR, our
calculations indicate that using the SDSS filters, the $\rm{MgII \,\, 2798 \,
\AA}$ line should be detected upto a redshift of 0.67 and the $\rm{ H\beta \,\,
4861 \, \AA}$ line upto 0.27 for a $10^6 M_{\odot}$ black hole. This is not
what is observed. On the other hand, when limits on the size of the BELR are
imposed, a $10^6 M_{\odot}$ black hole cannot be detected (for both $v_{min} =
1000 \,\, \rm{and} \,\,\kms$) using either of $\rm{MgII \,\, 2798 \, \AA}$ or
$\rm{ H\beta \,\, 4861 \, \AA}$ lines, a situation more in line with the
observations. 

The line ratios $\rm{MgII} / \rm{H\beta}$ vs $\rm{MgII} / \rm{CIV}$, for the
size limited BELR, using Keplerian cut-offs, are plotted using the solid red
lines in \fig{LineRatios}. The profiles are very different from the case of the
unrestricted BELR (dotted black lines). For a $10^8 M_{\odot}$, the values for
$\rm{MgII} / \rm{H\beta}$ are almost same for the size limited and the
unrestricted BELR, but the line ratio $\rm{MgII} / \rm{CIV}$ is very different
in the two cases. This line ratio varies from the unrestricted BELR case by
0.25 dex for $v_{min} = 1000 \kms$ and by 0.74 dex for $v_{min} = 3000 \kms$.
For a $10^6 M_{\odot}$, black, even the $\rm{MgII} / \rm{H\beta}$ ratio varies
from the unrestricted BELR case by 0.47 dex in the $v_{min} = 3000 \kms$
limited BELR. The $\rm{MgII} / \rm{CIV}$ line ratio varies from the
unrestricted BELR case by 0.59 dex for $v_{min} = 1000 \kms$ and by 1.24 dex
for $v_{min} = 3000 \kms$. Such variations in the line ratios for different
physical scenarios of the BELR, would act as diagnostics in our future
publications when we shall compare our theoretical results with observations
from SDSS or the likes of it, not only for these three emission lines, but for
many more appropriately chosen broad emission lines.


\subsubsection{Gravitational disk instability}
\labsubsubsecn{subsubsec:instability}

The outer parts of the $\alpha$-disk becomes self-gravitating and breaks up.
This radius has been suggested as the outer boundary of the BELR and the
beginning of the ``torus'' \citep{suganuma06, netzer93}. The radius at
which the disk becomes Toomre unstable 
\citep[the Toomre parameter $Q<1$][]{binney87} depends weekly on the
central black hole mass. Translating the Toomre criterion in terms of the
$\alpha$-disk parameters, $\Sigma/H = M_{BH}/R^3_{Instability}$, ($\Sigma$ is
the surface density and $H$ is the scale height of the disk) gives us
\begin{equation}  
R_{Instability}^{-\frac{9}{8}} = 1.31 \times 10^21 \alpha_{disk}^{} \left(\frac{\dot{m}} {\dot{m}_{\rm Edd}}\right)^{\frac{11}{20}} \left(\frac{M_{BH}}{M_{\odot}}\right)^{\frac{7}{40}} f^{\frac{11}{5}}
\labequn{equn:GravityUnstable}
\end{equation}  
where $f^4 = 1- \left(\frac{6GM_{BH}}{M_{\odot}}\right)^{\frac{1}{2}}$. The
much weaker mass evolution of $R_{Instability}$ (compared to $R_{Keplerian}$)
is shown as a dashed blue line in \fig{Rmax}. 

Using the $R_{Instability}$ cut off, below $M_{BH} \sim 10^{6.0} M_{\odot}$,
the $\mathcal{EW}$ distribution is same as when no radius restriction is
applied, leading to the unobserved increase in $\mathcal{EW}$ with decreasing
mass (dashed blue curves in \fig{AvgEwRcut}). The line corresponding line
ratios $\rm{MgII} / \rm{H\beta}$ vs $\rm{MgII} / \rm{CIV}$ are shown by the
dashed blue line in \fig{LineRatios}, further showing that the line fluxes are
exactly the same as that for an unrestricted BELR for $M_{BH} <= 10^{6.0}
M_{\odot}$

\begin{figure}
\begin{center}
\includegraphics[scale = 0.5, width = 8 cm, trim = 0 130 0 50, clip, angle = 0]{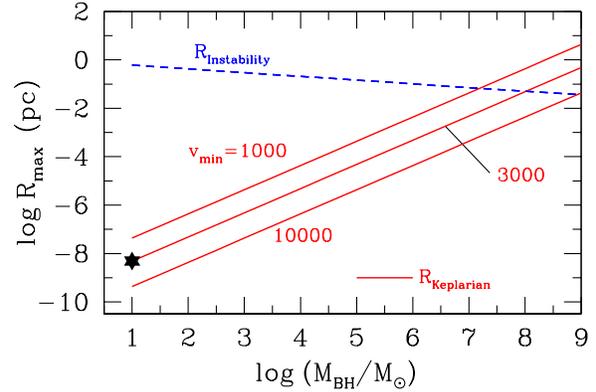}
\caption{The outer radius of BELR as a function of black hole mass. Various
methods of determining this cut-off radius are demonstrated : considering (a) a
phenomenological approach based on the observed velocities of the emission
lines (solid red lines) and (b) gravitational instability of the accretion disk
(dashed blue line). For stellar mass black hole ($M_{BH} = 10 M_{\odot}$), the
black star shows the tidal disruption radius.} 
\labfig{Rmax}
\end{center}
\end{figure}


\subsubsection{Tidal disruption by stellar mass black holes}
\labsubsubsecn{subsubsec:tidal}

According to \citet{hills75, gezari03}, 
\begin{equation}
R_{Tidal} = 1.1 R_{sch} \left(M_{BH}/10^8M_{\odot}\right)^{-2/3},
\labequn{Rtidal}
\end{equation}
for solar mass disrupted stars. Thus the disruption does not happen for $M_{BH}
> 1.1 \times 10^8M_{\odot}$ -- the stars are swallowed whole. Thus we use
\equn{Rtidal} to calculate $R_{Tidal}$ for a black hole of $M_{BH} = 10
M_{\odot}$ having a solar mass binary companion. The tidal disruption is
plotted as a black star in \fig{Rmax} and we find that $R_{Tidal} \sim
R_{Keplerian}$ for $v_{min} = 3000$. Our calculations for $\mathcal{EW}$
suggest zero flux in the emission lines for such low mass black holes (see
\fig{AvgEwRcut}) when the BELR size limit is imposed with $v_{min} = 3000
\kms$. No high sensitivity search for these emission lines have been conducted
for stellar mass black holes to corroborate or contradict these theoretical
results.  



\section{Discussion and Future Work}
\labsecn{sec:discussion}

\begin{enumerate}

\item[$\bullet$] Additional components of the AGN SED : \\
Both the AGN (super-massive black holes) and the XRBs (solar mass black holes)
are powered by the active matter accretion into the black hole. However, the
emission from the accretion disk is not the only component of the observed SED
from the AGN or the XRBs. In both cases we also observe power-law emission at
higher energies ($\geq 500 \ev$ for AGN and $\geq 3 \kev$ for XRBs). The
power-law is thought to be due to inverse comptonisation of some of the disk
photons by the hot coronal plasma surrounding the black hole, or due to the jet
(particularly for XRBs). In case of AGN of type 1, sometimes we see an
additional component, the soft-excess at $< 1 \kev$, which may be comptonized
disk emission or an entirely separate component altogether. Often the soft
excess can be modeled as a blackbody of temperature $100 - 200 \kev$. However,
for all the calculations presented in this paper, we ignore the power-law and
the soft-excess components of the SED, because their shape or strength are
independent of the mass of the black hole.  

The strengths of the lines are determined by the available number of photons
close to the ionization potential (IP) of the relevant ions. In this paper we
have restricted our studies to black holes with $M_{BH} \lesssim 10^9
M_{\odot}$ because it was sufficient to consider only the mass dependent
accretion disk component of the SED. In this mass range, the accretion disk is
hot enough that the energy ranges of the ionization or excitation potentials
($\lesssim 50 \ev$) required to produce  the lines $\rm{ H\beta \,\, 4861 \,
\AA}$, $\rm{CIV \,\, 1549 \,\AA}$ and $\rm{MgII \,\, 2798 \, \AA}$ are
dominated by the disk emission and the power-law or the soft excess would
contribute less than 1\% of the radiation from the accretion disk. 

To be thorough, we investigated the effect of adding a X-ray power-law with
spectral index $\alpha = 0.8$ and $\alpha_{OX} = 1.2$ for $M_{BH}\,=\, 10^{8.0}
M_{\odot}$. In each panel of \fig{AvgEwRcut}, the resultant $\mathcal{EW}$s are
shown as filled black triangles, for an unrestricted BELR, as filled red circle
for the $v_{min} = 3000 \kms$ Keplerian limit on the BELR and as filled blue
squares for the BELR restricted by gravitational instability. In each case
these points lie exactly on the line for the $\mathcal{EW}$ distribution,
generated using only the accretion disk component, showing that the addition of
the power-law and the soft-excess component is redundant for these emission
lines in the black hole mass range considered.

However, while considering similar studies for higher mass black holes, one has
to carefully account for the other aforementioned components of the AGN SED
because, the high energy tail of the accretion disk SED for $M_{BH} >= 10^9
M_{\odot}$ may cut-off at less than $50 \ev$.  

Similar considerations are required for constructing the AGN SED required to
correctly predict the $\mathcal{EW}$ for $\rm{OVI \,\, 1034 \, \AA}$, which is
an important BEL in the UV. The IP of OV is 113.90 eV, an
energy range dominated by the power-law for quasar-like ($M_{BH} \gtrsim
10^{8.0} M_{\odot}$) SED. Thus $\mathcal{EW}$ calculations for OVI would
require an AGN SED including all the components, and not just the accretion
disk emission. We will study such emission lines in details in our future
publications.


\item[$\bullet$] Alternative models for the accretion disk : \\
Instead of a radiatively efficient thin accretion disk \citep[BBB, due to][,
SS73]{shakura73}, sometimes black holes might have advection dominated,
radiatively inefficient accretion flows, which would result in significantly
different SEDs \citep[see e.g.][]{hopkins09} that the SS73 models (considered
in this paper). These alternative prescriptions drastically change the shape
of the SED in the energy range of the IP of the ions responsible for the BELs
and the narrow emission lines (NELs). We intend to calculate the line strengths
due to such ionizing SEDs and predict observable signatures, which might act as
diagnostic tools. 

Even for the BBB, more rigorous models exist for modeling the radiation from
the accretion disk (BBB). For example, \citet{blaes01, hubeny00, hubeny01,
hui05} discuss the role of real radiative transfer in the accretion disc. The
spin of the black hole is another physical parameter parameter to be considered
\citet{davis05, davis06}. Qualitatively speaking, for the same black hole mass
and accretion rate, these models push the peak of the SED to higher energies
than due to the SS73 model. We would like to calculate line strengths
corresponding to these models and test if the observations of BELs and NELs are
sensitive enough to differentiate these accretion disk models from BBB.


\item[$\bullet$] Comparison with data : \\ 
We intend to use the SDSS data base for $\rm{ H\beta \,\, 4861 \, \AA}$,
$\rm{CIV \,\, 1549 \,\AA}$ and $\rm{MgII \,\, 2798 \, \AA}$ and the HST and/or
FUSE data base for $\rm{OVI \,\, 1034 \, \AA}$, to compare our predicted
$\mathcal{EW}$ and line ratios with the observed line strengths. We would hope
to draw constraints on other physical parameters (e.g. distance of the BELR) by
such comparisons. Eventually we would want to extend our theoretical analysis
to include other fundamental black hole parameters like the accretion rate. We
would like to test if such systematic LOC modeling can explain observed effects
like the ``Balwin Effect'' \citep{baldwin77}, where the equivalent width of the CIV emission line decreases with increasing continuum luminosity.


\item[$\bullet$] Narrow Emission Lines : \\
We intend to do similar studies for the NELs in the AGN spectrum. It would be
interesting to see if they also show a mass dependence of the $\mathcal{EW}$,
given that the NEL clouds are further away from the black hole and may be
outside its sphere of influence (FWHM for NEL 1/6 of that for BELs) and/or
their sizes are not governed by Keplerian dynamics. 


\end{enumerate}


\section{Conclusions}
\labsecn{sec:conclusion}

\begin{enumerate}
\item[$\bullet$] We wanted to investigate if there is a lower mass cut-off
below which black holes cannot produce the broad emission lines (BELs) like $\rm{ H\beta
\,\, 4861 \, \AA}$, $\rm{CIV \,\, 1549 \,\AA}$ and $\rm{MgII \,\, 2798 \, \AA}$
typically used to detect and identify AGN activity from large optical data
bases like the Sloan Digital Sky Survey.  
\item[$\bullet$] Using the standard LOC (locally optically emitting clouds)
model for the broad emission line region (BELR) without any restrictions on the
radius on its radius, photoionization calculations show an unobserved rise in
the equivalent widths ($\mathcal{EW}$s) of the lines with decreasing mass.
\item[$\bullet$] However, introducing a cut-off radius for the BELR produces
sharp mass dependent drops in the $\mathcal{EW}$s, when the cut-off radius is
determined from simple Keplerian mechanics, depending on the mass of the black
hole ($M_{BH}$). Such drops are consistent with the observations including that
below $M_{BH} = 10^5 M_{\odot}$ the above mentioned emission lines are not
observed. Our findings conclude that these observations may not indicate the
absence of black holes of such low mass, but the inability of such black holes
to produce the observable (with our detectors) line strengths.
\item[$\bullet$] Such a conclusion might have consequences for modifying the black hole mass function in the lower mass end. However, before we can attempt to address such issues, we need to carry out a more rigorous systematic study (listed above, in \secn{sec:discussion}) of the $\mathcal{EW}$ of BELs and NELs, as a function of other fundamental black hole parameters like its accretion rate and alternative accretion theory models.
\end{enumerate}



\section{acknowledgments}
We thank Aneta Siemiginowska, Nirupam Roy and Yue Shen for helpful discussions and tips. We gratefully acknowledge the use of the Cosmology Calculator \citep{wright06}.



\end{document}